1

# Nontrivial Fixed Point in the 4D $\Phi^4$ Lattice Model with Internal $O(N)$ Symmetry

Julius Kuti [a] *

[a]Department of Physics 0319, University of California at San Diego, 9500 Gilman Drive, La Jolla, CA 92093-0319, USA

It is shown that the infinite dimensional critical surface of general euclidean lattice actions in a generic four-dimensional scalar field theory with $\Phi^4$ interactions has a domain of special multicritical points where higher dimensional operators play a special role. Renormalized trajectories of higher derivative continuum field theories with nontrivial interactions are traced back to special ultraviolet stable fixed points on the manifold of multicritical points. These fixed points have an infinite number of relevant directions which identify the universality classes of critical higher derivative field theories. The relevance of the new fixed point structure is discussed within the context of the triviality Higgs mass bound.

## 1. Introduction

It is generally accepted that the only fixed point on the infinite dimensional critical surface of a scalar lattice field theory with $\Phi^4$ interaction in four dimensions is an infrared stable Gaussian one. In this scenario, the only renormalized trajectory in the codimension of the critical surface, which in Wilson's terminology [1] defines a massive continuum scalar field theory, originates from the Gaussian fixed point when traced back to the critical surface, thereby describing a noninteracting system. Since the Gaussian fixed point is infrared stable, the massless scalar theory on the critical surface is also trivial. This picture would suggest that the critical surface of the $\Phi^4$ theory with internal O(N) symmetry is a manifold of simple critical points separating an ordered (ferromagnetic, or broken symmetry) phase from a disordered (symmetric) phase. A trivial renormalized trajectory runs into both phases from the Gaussian fixed point.

However, a scalar field theory with $\Phi^4$ interactions and higher derivatives in the quadratic part of the Lagrangian creates a seemingly paradoxical situation. It is a continuum field theory in four dimensions with nontrivial interactions [2]. In fact, a higher derivative term $\frac{1}{M^4}\Phi\Box^3\Phi$, with a new mass scale $M$ in the Lagrangian, renders the theory free of divergences. Since finite fluctuations cannot completely screen the $\Phi^4$ coupling, renormalized interactions will not vanish in the continuum. These theories cannot be described by the trivial renormalized trajectories of the infrared stable Gaussian fixed point.

One might suggest that $M$ now plays the role of the cut-off, therefore nothing has changed in the triviality scenario. This, however, cannot be a satisfactory resolution of the puzzle. It is legitimate to investigate the critical properties of higher derivative field theories when $M$ is kept on the scale of the low energy spectrum and cannot directly play the role of the cutoff. This happens when $M$ is kept fixed in inverse correlation length units of ordinary massive particles as the critical surface is approached ($\lim_{\xi\to\infty} M/\xi^{-1} = $ const) and $M$ becomes part of the low energy spectrum. By pushing the ratio $M/\xi^{-1}$ higher and higher, $M$ could ultimately take over the role of a regulator under some special circumstances. However, one can always make the fluctuations divergent again by adding new interaction terms, and the puzzle will reappear.

An explanation of the paradox will be offered by showing that the critical surface of a generic scalar lattice theory is more complex than suggested by the triviality scenario.

---

There exists a domain of special multicritical points where higher dimensional operators play a special role. The connection with the triviality picture which dominates the largest part of the critical surface can be explained by a complex crossover mechanism. The concept of renormalization group flows will guide our discussions.

## 2. Renormalization Group and Triviality

We will apply Wilson's renormalization group analysis [1] to general effective Hamiltonians of scalar field theories with a physical cutoff at a very large energy scale $\Lambda$. The theory at energies above $\Lambda$ could be a different field theory, or something completely new. The focus of our interest is the low energy description, far below the cutoff scale. Below the cutoff $\Lambda$, the most general effective Hamiltonian (euclidean action) in d dimensions is defined in terms of an infinite number of momentum dependent couplings $u^{(n)}(\vec{q}_1, \ldots, \vec{q}_n)$,

$$\mathcal{H} = \sum_{n=1}^{\infty} \frac{1}{n!} \int d^d q_1 \ldots d^d q_n \, u^{(n)}(\vec{q}_1, \ldots, \vec{q}_n) \\ \times \delta^{(d)}(\Sigma_i \vec{q}_i) \Phi(\vec{q}_1) \cdots \Phi(\vec{q}_n) \,. \quad (1)$$

The only constraints on the Hamiltonian come from the symmetries of the system. For example, only even terms are allowed in Eq. (1) under the $\Phi \to -\Phi$ Ising like symmetry. Although the implementation of the physical cutoff is quite arbitrary, we will assume a smooth momentum cutoff $\Lambda$ in continuous euclidean space, or a lattice discretization at small distances where the lattice spacing $a$ is related to the lattice momentum cutoff by $\Lambda = \pi/a$.

A renormalization group transformation involves the lowering of the cutoff from $\Lambda$ to $\Lambda_\rho$ and a rescaling of the field $\Phi$, as required by the critical properties of the system. The transformed effective Hamiltonian $\mathcal{H}_\rho(\Phi)$ is described by a set of new couplings $u_\rho^{(n)}(\vec{q}_1, \ldots, \vec{q}_n)$ which satisfy a coupled system of renormalization group equations. At a fixed point $\mathcal{H}^*$ the Hamiltonian is not changing, and the low energy theory becomes independent of the cutoff. The renormalization group flow of the effective Hamiltonian $\mathcal{H}_\rho$ around the fixed point is determined from a linear stability analysis. With $\mathcal{H}_\rho = \mathcal{H}^* + \Delta\mathcal{H}_\rho$ we write

$$\rho \frac{d}{d\rho} \Delta\mathcal{H}_\rho = L^*(\Delta\mathcal{H}_\rho) \,, \quad (2)$$

where $L^*$ is a linear operator. The solution of the linear renormalization group equations is given by

$$\Delta\mathcal{H}_\rho = \sum_i \mu_i(\rho) \mathcal{O}_i^* \,, \qquad \mu_i(\rho) = e^{y_i \rho} \,. \quad (3)$$

The eigenoperators $\mathcal{O}_i^*$ of $L^*$, with eigenvalues $\lambda_i = e^{y_i}$, are classified according to their scaling exponents $y_i$,

$$\begin{aligned} y_i &> 0 \quad \text{relevant eigenoperator } \mathcal{O}_i^* \,, \\ y_i &= 0 \quad \text{marginal eigenoperator } \mathcal{O}_i^* \,, \\ y_i &< 0 \quad \text{irrelevant eigenoperator } \mathcal{O}_i^* \,. \end{aligned}$$

Far away from the critical point, the expansion $\Delta\mathcal{H}_\rho = \Sigma_i \mu_i(\rho) \mathcal{O}_i^*$ is still applicable in terms of the complete set of eigenoperators $\mathcal{O}_i^*$. However, the $\rho$-dependence of the scaling fields $\mu_i(\rho)$ will be governed by nonlinear renormalization group equations, replacing the simple exponential evolution.

A Hamiltonian $\mathcal{H}$ at $\rho = 0$ is critical, if it converges to a fixed point with infinite correlation length, $\lim_{\rho \to \infty} \mathcal{H}_\rho = \mathcal{H}^*$. The couplings of critical Hamiltonians define the critical surface. The type of the critical behavior depends on the number of symmetry conserving relevant operators. For a normal critical point one has one relevant symmetry conserving operator which determines the critical temperature in statistical physics, or the critical mass parameter in euclidean field theory. At a multicritical point one has several relevant symmetry conserving operators, and consequently, several conditions for criticality.

In order to explore the critical domain of the Gaussian fixed point, we start from a simple critical Hamiltonian,

$$\mathcal{H} = \frac{1}{2} \int \Phi(\vec{q}) u^{(2)}(\vec{q}) \Phi(-\vec{q}) d^d q \,, \\ u^{(2)}(\vec{q}) = \vec{q}^{\,2} + \mathcal{O}(\vec{q}^{\,4}) \,, \quad (4)$$

which flows to the Gaussian fixed point $u_*^{(2)}(\vec{q}) = \vec{q}^{\,2}$ in the limit of a sharp momentum cutoff implementation of the renormalization group (other RG transfromations imply more complicated fixed point functions.) With scaling dimension $d_\Phi = \frac{1}{2}(d-2)$ for the scalar field $\Phi$, the eigenfunctions $u_*^{(n)}(\vec{q}_i)$ are homogeneous polynomials of degree $2k$ in the momentum variables $\vec{q}_i$. The scaling exponents are given by $y_{n,k} = d - \frac{n}{2}(d-2) - 2k$. At d=4, the only relevant eigenoperator $\int d^4q \Phi(\vec{q})\Phi(-\vec{q})$ has the scaling exponent $y_{2,0} = 2$ and it describes the mass term. The only nonredundant marginal eigenoperator is given by $\int \Phi^4(x) d^4x$ with scaling exponent $y_{4,0} = 0$. From the quadratic equation $\rho \frac{d}{d\rho}\mu_{4,0}(\rho) \sim \mu_{4,0}^2(\rho)$ of the marginal direction, the solution $\mu_{4,0}(\rho) \sim 1/\ln\rho$ flows with slow logarithmic rate to the Gaussian fixed point. All other eigenoperators are irrelevant.

The triviality scenario assumes that no other fixed points exist on the critical surface suggesting a manifold of simple critical points separating the ordered phase from the disordered phase. Earlier analytic and numerical studies had been consistent with this picture [3-7]. We will show, however, that there exists a region of special multicritical points on the critical surface which describes a new class of nontrivial continuum euclidean field theories. This will be demonstrated by a block spin renormalization group analysis on a simple hypercubic lattice in d dimensions.

## 3. Fixed Points in the Multicritical Region

Consider first the quadratic lattice Hamiltonian

$$\mathcal{H} = \frac{1}{2}\sum_{\vec{r}}\sum_{\vec{n}} \kappa(\vec{r}) \cdot \Phi(\vec{r})\Phi(\vec{n}+\vec{r}) , \qquad (5)$$

where a scalar field $\Phi(\vec{n})$ is associated with each lattice point $\vec{n}$. The general hopping term $\kappa(\vec{r})$ is not restricted to nearest neighbor interactions. A large finite volume, with $V = L^4$, and periodic boundary conditions $\Phi(\vec{n} + L\hat{e}_m) = \Phi(\vec{n})$, $\kappa(\vec{r} + L\hat{e}_m) = \kappa(\vec{r})$ are assumed in the calculation. The block spin transformation, as schematically depicted in Fig. 1, defines new fields $\Phi'(\vec{n}')$ on blocked sites $\vec{n}'$ where each block has $2^d$ sites of the original lattice. The transformation function of the block transfor-

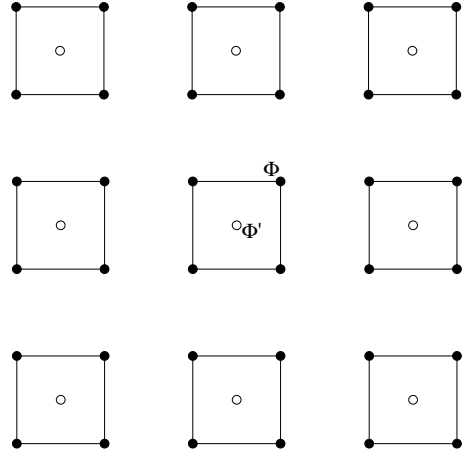

Figure 1. The blocked fields $\Phi'$ are defined on blocked sites $\vec{n}'$ which are designated by open circles.

mation is defined by [8]

$$\mathcal{T}_{a,b}[\Phi',\Phi] = -\frac{1}{2}a\sum_{\vec{n}'}\Big(\Phi'(\vec{n}') - b\sum_{\vec{j}\epsilon\vec{n}'}\Phi(\vec{j})\Big)^2 . \quad (6)$$

The parameter $a$ in Eq. (6) should not be confused with the lattice spacing which is set to one, so that the dimensional quantities are all measured in lattice spacing units. After k blocking steps the partition function $\mathcal{Z} = \int_\Phi e^{-\mathcal{H}(\Phi)}$ remains unchanged,

$$\mathcal{Z} = \text{const} \cdot \int_{\Phi'}\int_{\Phi} e^{\mathcal{T}_{a,b}^{(k)}[\Phi',\Phi] - \mathcal{H}(\Phi)} , \qquad (7)$$

where $\mathcal{T}_{a,b}^{(k)}[\Phi',\Phi]$ is obtained from Eq. (6) by the replacements $a \to a_k = \frac{a(1-2^db^2)}{1-(2^db^2)^k}$, and $b \to b_k = b^k$. The transformed Hamiltonian $\mathcal{H}^{(k)}(\Phi')$ is calculated by integrating out the variables $\Phi(\vec{n})$ of the original lattice,

$$\mathcal{Z} = \text{const} \cdot \int_{\Phi'} e^{-\mathcal{H}^{(k)}(\Phi')} , \qquad (8)$$

where a general quadratic form

$$\mathcal{H}^{(k)}(\Phi') = \frac{1}{2}\frac{1}{L'^4}\sum_{\vec{q}\,'} K^{(k)}(\vec{q}\,')\Phi'(\vec{q}\,')\Phi'(-\vec{q}\,') \quad (9)$$

is preserved. In Eq. (9) $L' = L/2^k$, and the Fourier transform $K(\vec{q}) = \sum_{\vec{r}} e^{-i\vec{q}\vec{r}}\kappa(\vec{r})$ is used in each blocking step.

With the choice $b = 2^{-(\frac{d}{2}+3)}$, the initial critical Hamiltonian with $K(\vec{q}) = z(\vec{q}^{\,2})^3 + \cdots$ will converge to a fixed point, $\lim_{k \to \infty} \mathcal{H}^{(k)} \to \mathcal{H}^*$, $\lim_{k \to \infty} K^{(k)}(\vec{q}) \to K^*(\vec{q})$, where

$$K^*(\vec{q}) = a^*\left(1 + \frac{a^*}{z}\sum_{\vec{l}=-\infty}^{+\infty}\frac{\prod_{i=1}^{d}\frac{\sin^2(q_i/2)}{(q_i/2+\pi l_i)^2}}{|\vec{q}+2\pi\vec{l}|^6}\right)^{-1}. \quad (10)$$

In Eq. (10) $a^* = a(1 - 2^d b^2)$ and $K^*(\vec{q})$ represents a line of fixed points parametrized by $a$. Higher powers $(\vec{q}^{\,2})^4, \ldots$ (irrelevant operators) in the initial couplings $K(\vec{q})$ of the critical Hamiltonian have no effects on the fixed point function $K^*(\vec{q})$. The normalization constant $z$ will have an important interpretation.

When $K^*(\vec{q})$ is expanded in powers of $\vec{q}^{\,2}$, the fixed point Hamiltonian in four dimesions has the simple form

$$\mathcal{H}^* \sim \int \left(z(\vec{q}^{\,2})^3 + \cdots\right)\Phi(\vec{q})\Phi(-\vec{q})d^4q \quad (11)$$

in the infinite volume limit, with eigenvalues $\lambda_{2,k} = 2^{6-2k}$ for the quadratic eigenoperators

$$\mathcal{O}_{2,k}^* \sim \int \left(\frac{K^*(\vec{q})}{z}\right)^2 (\vec{q}^{\,2})^{-6+k}\Phi(\vec{q})\Phi(-\vec{q}) . \quad (12)$$

A very similar picture is reproduced by using a differential renormalization group method in the continuum, if by analogy with the scale factor $b = 2^{-(\frac{d}{2}+3)}$, a new scaling dimension $d_\Phi = \frac{1}{2}(d-6)$ is chosen for the field $\Phi$. The scaling exponents, given by $y_{n,k} = d - \frac{n}{2}(d-6) - 2k$, identify three relevant quadratic operators at $n = 2$,

$$u_{2,0}^* \sim \int \Phi(\vec{q})\Phi(-\vec{q})d^4q , \qquad y_{2,0} = 6 ,$$

$$u_{2,1}^* \sim \int \vec{q}^{\,2}\Phi(\vec{q})\Phi(-\vec{q})d^4q , \qquad y_{2,1} = 4 ,$$

$$u_{2,2}^* \sim \int (\vec{q}^{\,2})^2\Phi(\vec{q})\Phi(-\vec{q})d^4q , \quad y_{2,2} = 2 .$$

One of the relevant quartic eigenoperators at $n = 4$ describes the $\lambda \int \Phi^4(x)d^4x$ interaction term in a general Hamiltonian,

$$u_{4,0}^* \sim \int \Phi^4(x)d^4x , \qquad y_{4,0} = 8 . \quad (13)$$

In fact, $y_{4,k}$ is positive for $k = 0, 1, 2, 3$. For general n, there exist an infinite number of relevant eigenoperators with positive scaling exponents. Their significance can be identified as new interactions which remain renormalizable by power counting in the presence of higher derivative propagators.

The extension of the above analysis to O(N) internal symmetry is straightforward. It is also important to note that an infinite class of new fixed points $K_\sigma^*(\vec{q}) = z(\vec{q}^{\,2})^\sigma + \cdots$ can be generated with positive integers $\sigma \geq 2$, if $b = 2^{-(\frac{d}{2}+\sigma)}$ and new scaling dimension $d_\Phi = \frac{1}{2}(d-2\sigma)$ are chosen. Our most studied model has $\sigma = 3$ which serves as an example in the discussion.

The newly found fixed points, as described by $K_\sigma^*(\vec{q})$, are generalized ultraviolet stable Gaussian fixed points in field theory terminology, with an infinite number of relevant directions. We will show now that the renormalized trajectories of higher derivative continuum field theories with nontrivial interactions are traced back to these special ultraviolet stable fixed points on the manifold of multicritical points.

## 4. Continuum Higher Derivative Theory

Consider the higher derivative Lagrangian

$$\mathcal{L} = -\frac{1}{2}\Phi_\alpha(x)\left(m^2 + \Box + M^{-4}\Box^3\right)\Phi_\alpha(x)$$
$$- \lambda_0\left(\Phi_\alpha(x)\Phi_\alpha(x)\right)^2 , \quad (14)$$

with internal O(N) symmetry, where a summation is understood over $\alpha = 1, 2, \ldots, N$. From a conventional viewpoint, the higher derivative term $M^{-4}\Box^3$ acts in the inverse euclidean propagator as a Pauli-Villars regulator with mass parameter $M$. It represents a minimal modification of the continuum model when its euclidean path integral is rendered finite in four dimensions [2]. The euclidean propagator is equivalent to the sum of three simple poles where a complex conjugate pair of ghost poles is added to original massive particle.

Since this model was discussed before [2], only the connection with the new fixed point structure will be presented. First, the scalar field will be given a new scaling dimension in the euclidean functional integral by the transformation $\varphi_\alpha = m^{-2}\Phi_\alpha$ where $m$, identified as the inverse correlation length after renormalization, is measured in units of a large momentum cutoff $\Lambda$. With the notation $z = m^4/M^4$, the couplings of the euclidean Hamiltonian, when expressed in terms of rescaled fields $\varphi_\alpha$, can be written as $u^{(2)}(\vec{q}) = m^6 + m^4 \vec{q}^{\,2} + z \cdot (\vec{q}^{\,2})^3$ and $u^{(4)} = \lambda_0 \cdot m^8$.

In the limit $m \to 0$, when $z$ is kept fixed, the critical surface is approached in the universality domain of the fixed point $K_3^*(\vec{q})$ which has three relevant quadratic operators in $\varphi_\alpha$. Therefore, a renormalized trajectory can be chosen which is traced back to

$$\mathcal{H}^* \sim \int z(\vec{q}^{\,2})^3 \varphi_\alpha(\vec{q}) \varphi_\alpha(-\vec{q}) d^4 q \qquad (15)$$

in the ultraviolet limit. It describes the continuum field theory of Eq. (14) with nontrivial interaction in the infrared region. The interaction term of Eq. (13) scales now as $\lambda_0 m^8 (\varphi_\alpha \varphi_\alpha)^2$ along the renormalized trajectory.

The nontrivial running coupling constant can be demonstrated by conventional renormalization methods providing a more complete understanding of the critical behavior [9].

## 5. Nontrivial Running Coupling Constant

The calculation of the scale dependent one-loop $\beta$-function in the broken phase will illustrate the nontrivial continuum limit of the theory. In addition to N-1 Goldstone excitations $\Phi_\alpha = \pi_\alpha$, $\alpha = 1, 2, \ldots, N-1$, there is also a massive Higgs excitation $\sigma$, $\Phi_N = \nu + \sigma$, where $\nu$ designates the renormalized vacuum expectation value of $\Phi_N$. A massive complex conjugate ghost pair appears in all channels, as a consequence of the higher derivative term in Eq. (14). The following scale dependent prescriptions specify the renormalization conditions.

1. The renormalization prescription on the one-particle irreducible Goldstone 2-point function, $\frac{d}{dp^2}\Gamma^{\pi\pi}(p^2)|_{p^2=0} = 1$, sets the correct normalization for the low energy Goldstone modes.

2. $(\frac{d}{dp^2})^3 \Gamma^{\pi\pi}(p^2)|_{p^2=M^2} = M^{-4}$ defines the renormalized ghost pole parameter $M$.

3. A scale dependent renormalized Higgs mass $m(\mu)$ is defined by $\Gamma^{\sigma\sigma}(\mu^2) = Z_1 \mu^2 + Z_3 \mu^6/M^4 + m^2(\mu)$ in the one-particle irreducible Higgs 2-point function. $Z_1$ and $Z_3$ designate renormalization constants in the counterterms of the derivative parts of the euclidean Hamiltonian.

4. The renormalization condition $\delta\nu = 0$ keeps the tree level relation $\nu = \frac{m^2}{8\lambda^2}$ exact in every order. This defines a scale dependent renormalized coupling constant $\lambda(\mu)$.

The scale dependent one-loop $\beta$-function is depicted in Fig. 2. At low energies the $\beta$-function is dominated by the Goldstone modes whose one-loop contribution is $\frac{N-1}{2\pi^2}\lambda^2(t)$. Above the Higgs mass threshold the Higgs loop kicks in and the $\beta$-function becomes $\frac{N+8}{2\pi^2}\lambda^2(t)$ which is the familiar one-loop form in the minimal mass independent subtraction scheme. The onset of ghost effects becomes dramatic at the scale $\mu = M^2$ where the $\beta$-function turns around and begins to drop rapidly towards zero as the Goldstone and Higgs loop contributions are being cancelled by ghost loops. The freeze of the running coupling constant above the ghost scale is shown in Fig. 3 indicating that the system remains weakly coupled in the entire energy range and the one-loop calculation is reliable.

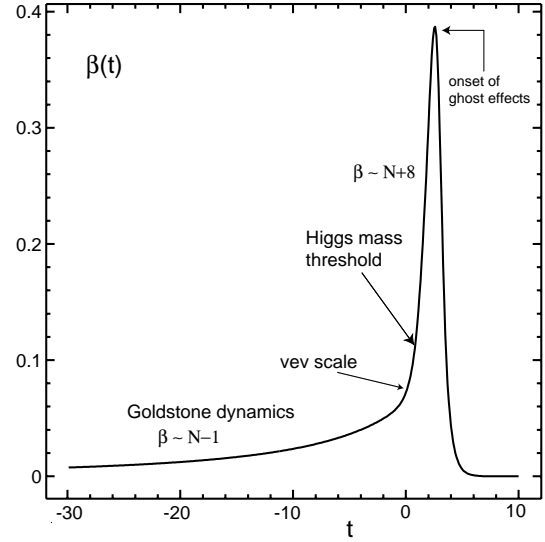

Figure 2. The one-loop $\beta$-function is plotted against the logarithmic scale $t = \ln\mu^2/\nu^2$ from the numerical integration of the renormalization group equations. The point $t = 0$ corresponds to $\mu = \nu$ and $t = 0$ on the logarithmic scale. The initial condition $\lambda(0) = 10/24$ is chosen with $M/\nu = 14.3$ which puts the ghost location into the multi-TeV range.

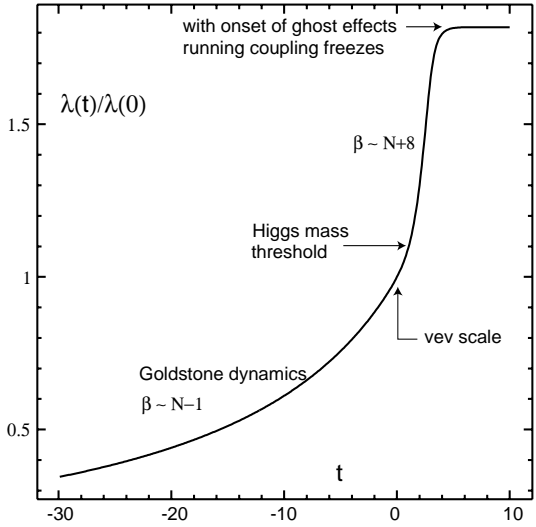

Figure 3. The running coupling constant is plotted from the numerical integration of the one-loop renormalization group equations with input parameters as defined earlier.

## 6. Fate of the Lattice Higgs Mass Bound

A hypercubic lattice structure was introduced [2,10,11] in non-perturbative computer simulations of the higher derivative Lagrangian of Eq. (14). The phase diagram is shown in Fig. 4 in the limit of infinite bare coupling. Tuning the hopping parameter $\kappa$ to the critical line for fixed $M$ corresponds to the triviality limit of the scalar field theory. In this limit, the dimension eight operator $M^{-4}\Phi_\alpha\square^3\Phi_\alpha$ becomes irrelevant and the dotted critical line with the exception of the end point at the origin represents a single universality class. The continuum limit of the higher derivative theory without the underlying lattice structure is equivalent to the tuning of $\kappa$ towards the origin along a line of fixed $z$. A multicritical point is approached in this limit, as indicated by the solid line of Fig. 4. The higher derivative term $M^{-4}\Phi_\alpha\square^3\Phi_\alpha$ becomes a relevant operator in this limit, and the theory is governed by the nontrivial fixed point.

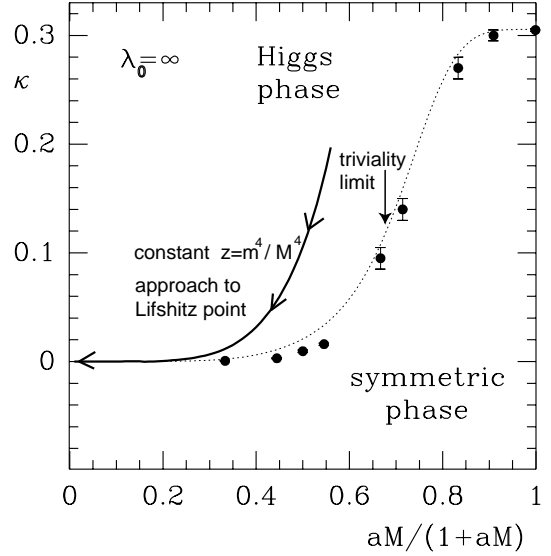

Figure 4. The phase diagram of the lattice model at infinite bare coupling. The dotted line is calculated in the large-N expansion. The solid line displays the fixed $z = m^4/M^4$ ratio towards the continuum limit of the higher derivative theory.

Higgs mass values from lattice simulations are shown in Fig. 5. In the 600 GeV to 700 GeV range, the solid line of the simple hypercubic lattice action does not include any higher dimensional operators [3–5]. The dashed line corresponds to an F4 lattice action with dimension six *irrelevant* operators [7]. A Symanzik improved simple cubic lattice action is also shown with dimension six *irrelevant* operators towards the restoration of euclidean invariance at finite correlation length [6]. In the much higher Higgs mass range, between 1.6 TeV and 1.8 TeV, a dimension eight term is added as a *relevant* operator to the euclidean action on a simple cubic lattice, to describe a strongly interacting Higgs sector with higher derivative Lagrangian [2,10,11]. The ghost location in this case is placed in the multi TeV mass range, and the lattice spacing could be completely eliminated.

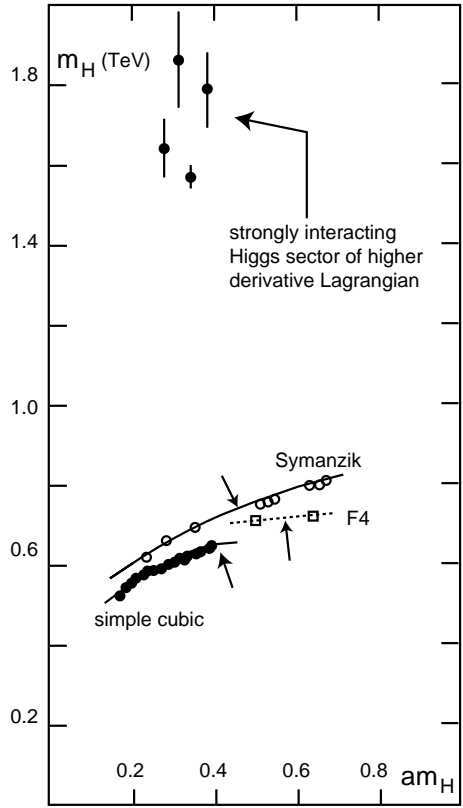

Figure 5. The Higgs mass bound of several lattice simulations is depicted. The arrows indicate one percent deviation from euclidean invariance in a scattering amplitude as defined in [4].

Large $N$ calculations also indicate that at $m_H$ = 700 GeV the ratio $M/m_H$ is of the order of 30 at infinite bare coupling. With the complex ghost location in the 20 TeV mass range, the $M$ dependence in scattering amplitudes becomes practically invisible, well within the intrinsic ambiguity of the perturbative expansion. This is in sharp contrast with earlier lattice models [3–7] where the violation of euclidean invariance was larger than the ambiguity of the perturbative expansion in the $m_H$ = 700 GeV mass range. We have to suggest, therefore, that the Higgs mass bound in the 700 GeV mass range in the earlier calculations was imposed by the artifacts of the underlying lattice structure.

Since the complex conjugate ghost pair of the higher derivative Lagrangian theory evades easy experimental detection without violating unitarity [9,12], or Lorentz invariance, it serves as a model of the strongly interacting Higgs sector without technicolor, a scenario which was excluded in previous lattice studies.

## Acknowledgements

This work was supported by the DOE under Grant DE-FG03-90ER40546. I am grateful to my collaborators Chuan Liu and Karl Jansen who allowed me to report here some of the simulation results on the Higgs mass bound and the large-N analysis of unitarity prior to publication. The simulations were done at Livermore National Laboratory with DOE support for supercomputer resources.